\documentclass[epj]{svjour}
\usepackage{graphics}
\usepackage{times}
\usepackage{helvet}
\begin{document}
\title{Hydrodynamic description of spectra at high transverse mass in 
ultrarelativistic heavy ion collisions}
\author{T.~Peitzmann}
\institute{Universiteit Utrecht, NL-3508 TA Utrecht, The Netherlands}
%
%
\abstract{
Transverse mass spectra of 
pions and protons measured in central collisions of heavy ions at 
the SPS and at RHIC are compared to a hydrodynamic parameterization. 
While the chemical temperature needed at RHIC is significantly higher 
compared to SPS, the spectra may be described using kinetic 
freeze-out parameters which are similar for both beam energies. 
At RHIC either the temperature or the flow velocity is 
higher, but the data provide no unambiguous proof for 
much stronger transverse flow. 
The contribution of such hydrodynamic emission at high transverse 
momenta is investigated in detail. It is shown that hydrodynamics may 
be relevant up to relatively high transverse momenta. The importance 
of the velocity profile used in this context is highlighted. 
} 
\maketitle
\section{Introduction}

The description of transverse momentum (or transverse mass) spectra 
with hydrodynamic models is one of the well established tools in 
ultrarelativistic heavy-ion physics 
\cite{Schn93,Chap95,PBM95,wiedemann96,na49hydro,wa80:pi0:98,wa98:hydro:99,tomasik99}.
The applicability of such models is still under discussion, because 
they require local thermalization, which may not hold for the entire 
reaction volume. Furthermore, at high transverse momenta hard 
scattering is expected to provide an important contribution to 
particle production -- the exact momentum region where this would 
apply is under debate.

\begin{sloppypar}
Comparisons to experimental data have often been per\-formed 
using parameterizations of hydrodynamic distributions, which have to 
use simple assumptions about the freeze-out conditions, e.g. mostly 
one universal freeze-out temperature is used. While such 
parameterizations can not replace full hydrodynamic calculations, 
they can provide a reasonable guideline to the general behavior of 
particle distributions, especially as systematic studies are more 
accessible in this case due to the much more modest calculational 
effort.
\end{sloppypar}

In the present paper I will assume the validity of hydrodynamic 
pictures, but I will not attempt to deliver a full-fledged, 
realistic hydrodynamic calculation. A parameterization will 
be used to describe momentum spectra at low and intermediate 
$m_{T}$. \footnote{While such calculations can also be used to 
compare to two-particle-correlations, this will not be the scope of 
this paper.} Parameters for chemical and kinetic freeze-out will be 
determined from a simultaneous fit of pion and (anti-)proton spectra. 
The main intention is to use a not too unrealistic 
hydrodynamic parameterization fitted to low transverse masses to study 
the extrapolation to high $m_{T}$.

This is very important also in the light of recent attempts to measure jet 
quenching at RHIC via transverse momentum distributions \cite{highpt}. If the 
particle emission at low $m_{T}$ is governed by hydrodynamics, as 
e.g. the results on elliptic flow \cite{starv2id,phenixv2} suggest, 
then such a hydrodynamic source will inevitably contribute at high 
$m_{T}$ also. Such a hydrodynamic contribution would have to be 
taken into account before comparing them to 
pQCD calculations and obtaining estimates of the amount of 
suppression of hard scattering.

\section{The Model}

The calculations presented in this paper are based on a 
parameterization of the source at freeze-out by 
Wiedemann and Heinz \cite{wiedemann96} which is motivated by a
hydrodynamic approach and includes effects of transverse flow and 
resonance decays. The original computer program
calculates the direct production of pions and the 
contributions from the most important resonances having two- or three-body 
decays including pions ($\rho$, $\mathrm{K}^{0}_{S}$, 
$\mathrm{K}^{\star}$, $\Delta$, $\Sigma + \Lambda$, $\eta$, $\omega$, 
$\eta^{\prime}$).  
The transverse momentum spectra of the directly emitted resonances $r$ 
are given by: 
  \begin{eqnarray}
  \label{eq:urs1}
    {dN_r^{\rm dir}\over dM_T^2} 
    &=& \mathrm{const.} \cdot \, (2J_r+1) \, 
  \nonumber\\     
    &&\times \, M_T 
	\int_0^4 d \xi e^{-\xi^2/2}\,
        K_1\left( {\textstyle{M_T\over T}}\cosh\eta_t(\xi) \right)
  \nonumber \\
    && \times \,
    I_0\left( {\textstyle{P_T\over T}}\sinh\eta_t(\xi) \right) \, ,
  \end{eqnarray}
where $\xi = r/R$ and $R$ is the Gaussian radius of the source.
Because the integrals are evaluated numerically, fixed integration
boundaries have to be chosen. The upper limit of $\xi = 4$ was chosen
to ensure that for most reasonable shapes only negligible tails of
the distribution extend beyond the limits.
The transverse rapidity of the source element is given as 
$\eta_t(\xi)=\eta_f \xi^n$, where the default value of the power is 
$n=1$. The transverse rapidity parameter $\eta_{f}$ controls the amount 
of transverse flow.

This 
distribution represents the rapidity integrated spectrum, which will 
be used in this paper, as it is easier to calculate compared to the 
general rapidity differential spectrum. In comparison to 
experimentally measured spectra, which are mostly from limited 
rapidity regions, this may introduce a bias. This bias should however be 
small, if the rapidity dependence of the spectra is negligible as in a 
longitudinal scaling expansion (Bjorken) scenario. In fact, although 
the Bjorken picture is not at all applicable at SPS, the
dependence of the spectral shape on rapidity is already reasonably 
small \cite{na49jones}. The dependence of yields of different species
on rapidity may have a small effect on the chemical parameters in the fits.
We will comment on this again in a later section.

The original version of this model uses the following assumptions:
\begin{enumerate}
    \item  The spatial density distribution at freeze-out is chosen 
    as a Gaussian. 

    \item  One universal freeze-out temperature is used which 
    determines both the spectral shape and the ratio of different 
    particle species (i.e. resonances). 
\end{enumerate}

In the calculations performed here the model has been modified in the 
following way:
\begin{enumerate}
    \item  As the spatial distribution a Woods-Saxon shape:
    \begin{equation}\label{eq:woodssaxon}
	\rho_{\mathrm{ws}}(\xi) = \rho_{0} \cdot \frac{1}{1 + \exp{\Delta(\xi - 1)}}
    \end{equation}
    has been introduced. By varying the parameter $\Delta$ the shape 
    can be adjusted. For large values of $\Delta$ the shape approaches 
    a box-like distribution. The shape may also be chosen very similar 
    to a Gaussian for comparison. 
    The spatial distributions fix implicitly the shape of the 
    velocity distribution. While this may not be very relevant at low 
    transverse momenta, it has been shown to be important at high 
    $p_{T}$ \cite{wa98:hydro:99}. The influence of different 
    distributions has also been investigated in \cite{tomasik99}.

    \item  The spectra for a given particle (or resonance) are 
    calculated using the kinetic temperature $T_{kin}$. Finally the 
    normalization is readjusted to the chemical temperature $T_{chem}$ 
    assuming that $dN/dy$ at 
    midrapidity scales with the temperature as \cite{Schn93}:
    \begin{equation} \label{dndythermal}
        {{dn_{\rm th}}\over{dy}} = { V \over {(2\pi)^2} } T^3
	\Big(
	{{ m^2 }\over{T^2}}  + {2m\over T}+ 2 
	\Big)
 	\exp\left(-{m\over T} \right) \; .
    \end{equation}
    It is commonly believed that chemical freeze-out 
    (determining particle ratios) 
    should occur at higher temperature than 
    kinetic freeze-out (determining the spectral shape), 
    so at least two independent temperatures may be needed.

    \item  Furthermore the program has been enhanced to simultaneously 
    describe protons and antiprotons in addition to pions. Here the 
    decay contributions from $\Delta$ and $\Sigma + \Lambda$ have been 
    taken into account. This of 
    course requires the introduction of another parameter: the 
    baryonic chemical potential $\mu_{B}$.
\end{enumerate}

\begin{figure}[bt]
        \includegraphics{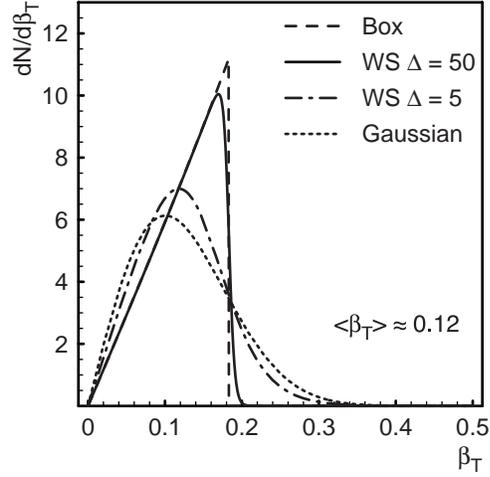}
        \caption{Velocity distributions of the model source for 
        different source shapes at an average 
        velocity of $\langle \beta_T \rangle \approx 0.12$.}
        \protect\label{fig:vel1}
\end{figure}

To better understand the importance of the different shapes of the 
source distribution, it is helpful to investigate the resulting 
velocity distributions:
\begin{equation} \label{eq:velocity}
    \frac{dN}{d\beta_T} = \frac{2 \pi \xi}{\eta_{f}}\cdot 
    \frac{1}{1-\left[ \beta_T(\xi) \right]^{2}}\cdot \rho(\xi),
\end{equation}
and their mean values:
\begin{equation} \label{eq:meanbeta}
    \langle \beta_T \rangle = \frac{\int_0^4 d\xi \beta_T(\xi) \rho(\xi)}
    {\int_0^4 d\xi \rho(\xi)},
\end{equation}
as they directly influence the shape of the momentum distribution, 
while the spatial distribution is actually not directly relevant.
Fig.~\ref{fig:vel1} shows velocity distributions for different shapes 
at a small average velocity of $\langle \beta_T \rangle \approx 0.12$. For 
such small velocities the relation between rapidity and velocity 
$\beta_T = \tanh \eta_{t}$ may be approximated by a linear relation, so most 
of the expected properties of the distributions are preserved.
One can see that the box profile yields a sharp cutoff of the 
velocity distribution. The Woods-Saxon with $\Delta = 50$ looks very 
similar to the box with a slightly smeared out edge. The Gaussian 
has a much more smooth edge resulting from the tail of 
the spatial distribution. The Woods-Saxon with $\Delta = 5$ provides 
a similar case as the Gaussian, both have considerable contributions 
at velocities much higher than the average which will finally lead to 
an inhanced yield at higher transverse momenta for the same inverse 
slope a low $p_{T}$, i.e. to stronger curvature of the spectrum.

\begin{figure}[bt]
        \includegraphics{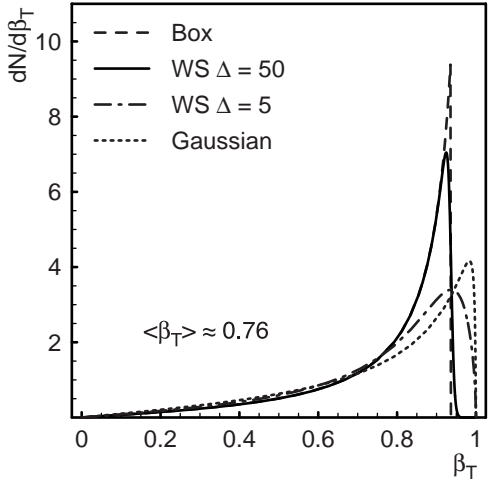}
        \caption{Velocity distributions as in Fig.~\protect\ref{fig:vel1} 
	at an average 
        velocity of $\langle \beta_T \rangle \approx 0.76$.}
        \protect\label{fig:vel2}
\end{figure}

Fig.~\ref{fig:vel2} shows similar distributions for a high average 
velocity of $\langle \beta_T \rangle \approx 0.76$. Again the box profile 
leads to a sharp upper limit in velocity, but the triangular shape of 
the distribution is distorted by the non-linearity of the relation 
between $\beta_T$ and $\eta_{t}$. The Woods-Saxon for $\Delta = 50$ is 
again similar to the box, and the one for $\Delta = 5$ and the Gaussian 
are also close. However, in this case the two latter distributions 
have a maximum shifted towards higher velocity which is also due to 
the non-linearity. Still, the general property of these distributions 
is that they have larger contributions at high velocities for the same 
average velocity as the other two distributions. For such large
velocities it is noteworthy that the average velocity depends
significantly on the recipe used to calculate it. If instead of
equation~\ref{eq:meanbeta} one uses:
\begin{equation} \label{eq:meanbeta1}
    \langle \beta_T \rangle^{\prime} = \frac{\int_0^4 d\xi \beta_T(\xi) 
    \gamma_T(\xi) \rho(\xi)}
    {\int_0^4 d\xi \gamma_T(\xi) \rho(\xi)},
\end{equation}
the numerical values are larger by almost $10 \%$, and differ
slightly for the different profiles. I will use the definition
\ref{eq:meanbeta} throughout.

\section{Hadron Spectra from SPS}

\begin{figure}[tb]
        \includegraphics{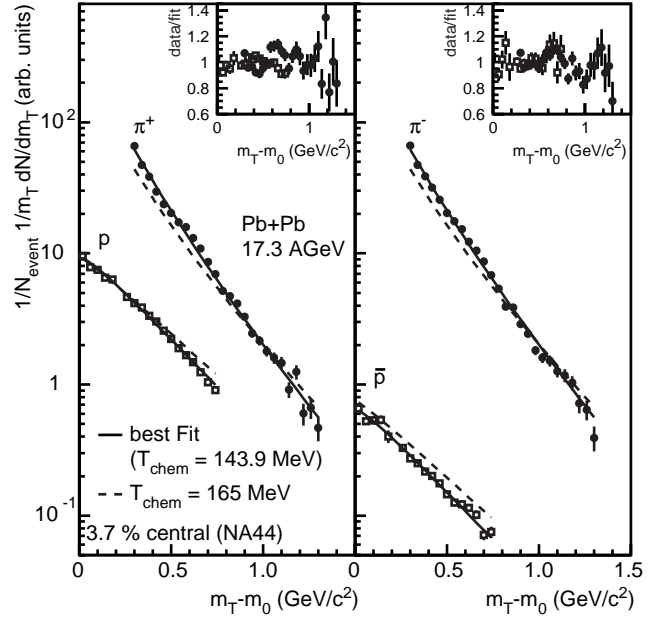}
        \caption{Transverse mass distributions of pions and 
        (anti-)protons for central Pb+Pb collisions at 158 $A$GeV 
        from the NA44 experiment \protect\cite{na44spec}. Included are 
        two different fits to the data with the hydrodynamic 
        model discussed in the text.}
        \protect\label{fig:na44}
\end{figure}

\begin{sloppypar}
The model described above has first been compared to data from SPS 
heavy ion experiments. Data on pion, proton and antiproton 
production in the 3.7 $\% $ most central 
Pb+Pb collisions from NA44 \cite{na44spec} and
in the 10 $\% $ most central 
Pb+Pb collisions from NA49 \cite{na49spec1,na49spec2}
have been used. The transverse mass spectra of NA44 
are shown in Fig.~\ref{fig:na44}, they cover the range of 
\mbox{$0.3 \, \mathrm{GeV}/c^{2} \le m_{T}-m_{0} \le 1.3 \, \mathrm{GeV}/c^{2}$} 
for pions and 
$0.02 \, \mathrm{GeV}/c^{2} \le m_{T}-m_{0} \le 0.74 \, 
\mathrm{GeV}/c^{2}$ for protons and antiprotons. The model has been 
fitted to the data. The best agreement can be achieved with a kinetic 
temperature $T_{kin} = 122.2 \, \mathrm{MeV}$, an average transverse 
flow velocity $\langle \beta_T \rangle = 0.478$, a chemical 
temperature $T_{chem} = 143.9 \, \mathrm{MeV}$ and a baryonic chemical 
potential $\mu_{B} = 193.3 \, \mathrm{MeV}$. For this fit the width 
parameter has been set to $\Delta \equiv 50$, allowing for a free 
variation of this parameter produces a slightly smaller value with a 
very large error while the other parameters remain unchanged. As the 
fits seem to be insensitive to small changes in $\Delta$, we have 
performed most fits with a fixed value. For systematic checks some of 
the constraints on the fits have been varied. Most of these changes, 
as e.g. omitting some of the data points (the lowest or highest in 
$m_{T}$, resp.) from the fit, do not affect the parameter values 
significantly. Significant changes are obtained under the following 
conditions:
\end{sloppypar}

\begin{figure}[tb]
	\includegraphics{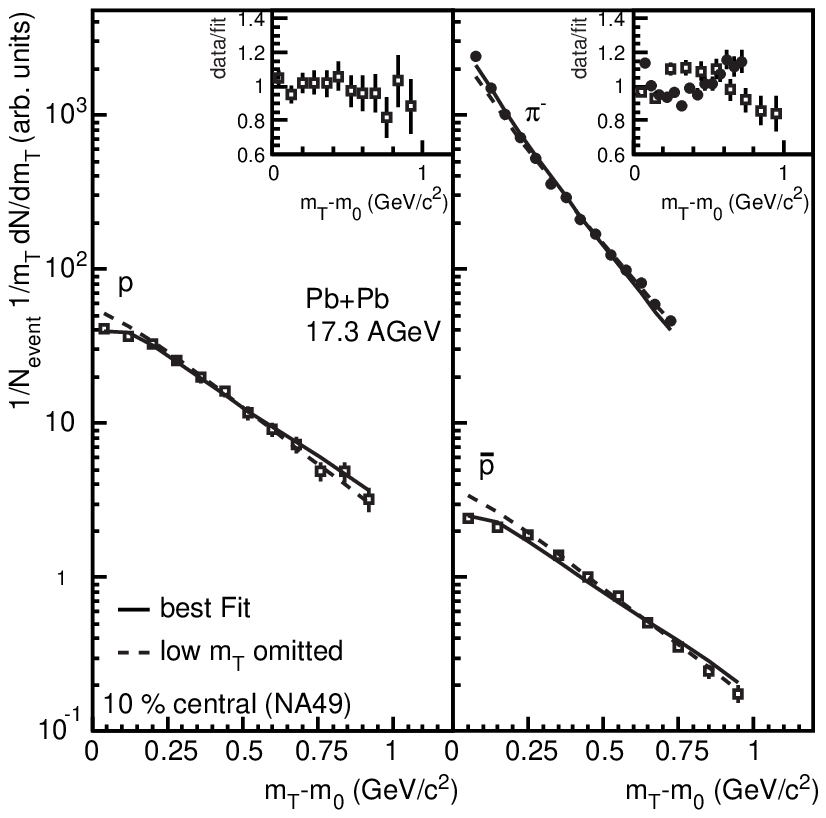}
	\caption{Transverse mass distributions of pions and 
	(anti-)protons for central Pb+Pb collisions at 158 $A$GeV 
	from the NA49 experiment \protect\cite{na49spec1,na49spec2}. 
	Included are 
	two different fits to the data with the hydrodynamic 
	model discussed in the text.}
	\protect\label{fig:na49}
\end{figure}

\begin{table*}[tbh]
    \centering
    \caption{Fit parameters of hydrodynamic fits to pion and 
    (anti-)proton spectra in central Pb+Pb collisions at the CERN 
    SPS. The errors given are statistical. No error indicates 
    that this parameter was fixed in the fit.}
    \begin{tabular}{|c||c|c|c|c|c|c|c|}
        \hline
        Remarks & $\chi^{2}/\nu$ & $T_{kin}$ (MeV) & $\eta_{f}$ 
	& $\langle \beta_T \rangle$ & $T_{chem}$ (MeV) 
	& $\mu_{B}$ (MeV) & $\Delta$  \\
        \hline \hline
	\multicolumn 8{|c|}{NA44 data} \\ 
	\hline
        best fit & $217/82$ & $122.2 \pm 1.9$ & $0.805 \pm 0.007$ 
	& $0.478 \pm 0.004$ & $143.9 \pm 0.6$ & $193.3 \pm 0.9$ & $50$  \\
        \hline
        no weak decays & $217/82$ & $120.4 \pm 1.9$ & $0.804 \pm 0.007$ 
	& $0.478 \pm 0.004$ & $150.9 \pm 0.4$ & $202.6 \pm 0.9$ & $50$  \\
        \hline
        broad profile & $360/82$ & $95.6 \pm 3.0$ & $0.745 \pm 0.010$ 
	& $0.517 \pm 0.007$ & $146.3 \pm 0.4$ & $196.4 \pm 0.9$ & $5$  \\
        \hline
        $T_{kin} = T_{chem}$ & $358/85$ & $144$ & $0.728 \pm 0.004$ 
	& $0.442 \pm 0.002$ & $144$ & $193.3$ & $50$  \\
        \hline
        high $\langle \beta_T \rangle$ & $799/83$ & $92.7 \pm 0.9$ & 
        $0.968 \pm 
        0$ & $0.55$ & $144.6 \pm 0.4$ & $193.2 \pm 0.9$ & 
        $50$  \\
        \hline
        high $T_{chem}$ & $2394/83$ & $156.3 \pm 3.0$ & $0.802 \pm 
        0.001$ & $0.477 \pm 0.001$ & $165$ & $208.4 \pm 0.8$ & 
        $50$  \\
	\hline \hline
	\multicolumn 8{|c|}{NA49 data} \\ 
	\hline
	best fit & $129/31$ & $75.8 \pm 1.9$ & $0.968 \pm 0.010$ 
	& $0.550 \pm 0.004$ & $133.2 \pm 0.5$ & $184.5 \pm 2.0$ & $50$  \\
	\hline
	low $m_{T}$ omitted & $16/22$ & $108.0 \pm 6.6$ & $0.828 \pm
	0.034$ 
	& $0.489 \pm 0.016$ & $136.6 \pm 0.9$ & $184.6 \pm 3.3$ & $50$  \\
	\hline
    \end{tabular}
    \label{tbl:parsps}
\end{table*}

\begin{enumerate}
    \item  All contributions from weakly decaying particles are 
    ignored. 
    
    \begin{sloppypar}
    In this case the chemical temperature and the baryonic 
    chemical potential change slightly, while the kinetic temperature 
    and the flow velocity are essentially unaltered. This is 
    understandable as the particle ratios may depend more strongly on 
    this contribution than the shape of the spectra. As the quality 
    of the fit is similar to the best fit above and a not 
    precisely specified fraction of weak decays may contribute to the 
    data, this has to be taken into account in the systematic error. 
    \end{sloppypar}

    \item  The width parameter is set to $\Delta \equiv 5$.
    
    In this case the chemical parameters show small changes, while the 
    temperature and flow velocity change more significantly. This is
    mainly due to the fact that this spatial distribution, which 
    is similar to a Gaussian, results in a broader velocity distribution. 
    The quality of the fit is significantly worse than the fits given 
    above.

    \item  The kinetic and chemical temperatures have been set to the 
    same value of $T \equiv 144 \, \mathrm{MeV}$ and the chemical 
    potential to $\mu_{B} = 193.3 \, \mathrm{MeV}$.
    
    Here, the flow velocity obtained is smaller than in the best fit, 
    as is expected for a higher kinetic temperature. Again the fit 
    quality is worse.

    \item  The expansion velocity has been fixed to a value of 
    $\langle \beta_T \rangle = 0.55$. The fit 
    quality is much worse than the best fit.

    \item  The chemical temperature has been set to $T_{chem} \equiv 
    165 \, \mathrm{MeV}$ similar to results from fits of hadrochemical models 
    to ratios of total multiplicities of different species \cite{pbm99}.
    
    This fit cannot describe the spectra, it essentially has the 
    wrong ratio of pions to protons. Of importance may be 
    the different integration region in rapidity compared to \cite{pbm99}. 
    If e.g. the rapidity distribution of protons 
    is broader than the one of pions this may easily explain the 
    smaller chemical temperature obtained here. However, for the purpose 
    of this paper, we will just note this discrepancy and use the fit 
    results as they are. The parameters obtained apparently describe 
    the particle ratios at midrapidity investigated here, and this 
    may be the relevant information which should be used also in 
    estimating the contribution of resonances to the momentum spectra 
    at this given rapidity. 
    
    Furthermore, it may be that part of the longitudinal motion of the 
    source is not generated hydrodynamically but is a remnant of the 
    initial state motion, which may be even more likely for 
    participant protons. In this case it is not completely obvious, 
    whether the protons observed at very different rapidities share 
    the same chemical freeze-out temperature. If the 
    spacetime-momentum correlation originates from a very early phase 
    of the collision, a unique chemical temperature might not be 
    applicable and the use of integrated multiplicities might be 
    misleading. Rather, there is the possibility of a local 
    temperature which may be applicable for limited rapidity regions.
\end{enumerate}
The fit parameters obtained from the fits discussed above are 
summarized in table~\ref{tbl:parsps}. None of the fits provides a 
perfect agreement with the data, which is due to a structure in the 
experimental spectra which can not be described within this model.

\begin{sloppypar}
The transverse mass spectra of NA49 cover the range of 
\mbox{$0.05 \, \mathrm{GeV}/c^{2} \le m_{T}-m_{0} \le 0.75 \,
\mathrm{GeV}/c^{2}$} for pions and 
\mbox{$0 \, \mathrm{GeV}/c^{2} \le m_{T}-m_{0} \le 1 \, 
\mathrm{GeV}/c^{2}$} for protons and antiprotons, 
they are shown in Fig.~\ref{fig:na49} together with fits of the model. 
As pions and protons are not available for the same
centrality selection -- pions have been measured for 5 $\% $ central 
and (anti-)protons for 10 $\% $ central -- the pions have been
rescaled by a factor of $1/1.08$ according to the number of participants 
in both samples \cite{marek}. This will introduce an additional
systematic error to the chemical parameters of the fits, which depend
most significantly on the relative normalization of different species.
The fitting conditions have been varied as above, 
the relative variation of the parameters is also similarly small. In fact,
the chemical parameters are relatively similar with slightly smaller 
values compared to the NA44 data.
However, for these data the best agreement can be achieved with a
very low kinetic temperature $T_{kin} = 75.8 \, \mathrm{MeV}$ and a
very high average transverse 
flow velocity $\langle \beta_T \rangle = 0.55$.  It turns out that
this significant difference to the NA44 data is to a large extent due
to the behavior of the data (particularly of the protons and
antiprotons) at very low $m_{T}$. If the lowest three data points of all spectra 
are omitted from the fit, a kinetic 
temperature $T_{kin} = 108.0 \, \mathrm{MeV}$ and an average transverse 
flow velocity $\langle \beta_T \rangle = 0.489$ are obtained with a 
much better $\chi^{2}$. The parameters of this fit are relatively
close to those obtained from the fit to NA44 data. The low $m_{T}$
data points of NA49 appear to disagree with the measurements by NA44.
It is also seen from the best fit to NA49 data that the agreement is 
not perfect if all data points are included -- there is a tendency of
a larger inverse slope of the fit compared to the data especially for 
the antiprotons.
\end{sloppypar}

\begin{figure}[tb]
        \includegraphics{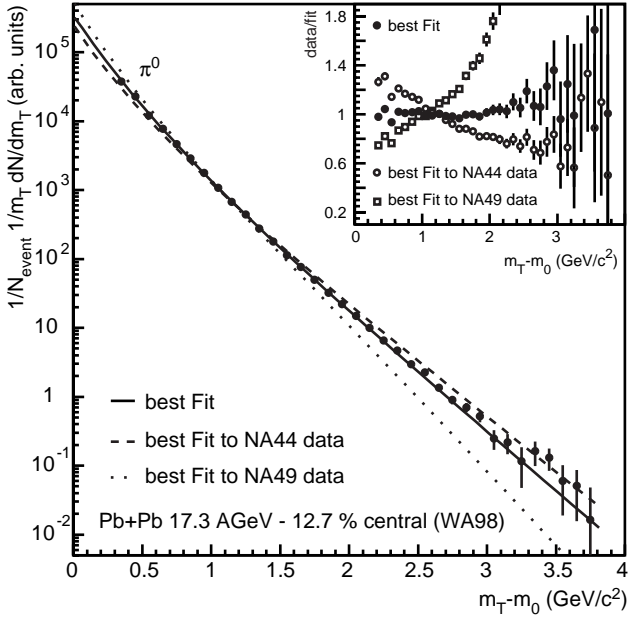}
        \caption{Transverse mass distributions of neutral pions for 
        the $12.7 \%$ most central Pb+Pb collisions at 158 $A$GeV 
        from the WA98 experiment \protect\cite{wa98pi0}. Included are 
        two different fits to the data with the hydrodynamic 
        model discussed in the text.}
        \protect\label{fig:wa98}
\end{figure}

While the data used above have a limited range in transverse mass, one 
might also consider comparing such hydrodynamic distributions to 
data with a wider transverse momentum range. This has actually been 
done before at the SPS by WA98 \cite{wa98:hydro:99}, using also a 
program derived from the one of Wiedemann and Heinz 
\cite{wiedemann96}. However in this version the original assumption 
of the same temperature both for chemical and kinetic freeze-out 
was applied. Also, only neutral pion spectra were used, which give 
very limited information to determine all relevant parameters.

In the following, information from the above fits will be 
used while fitting the neutral pion spectra of WA98 \cite{wa98pi0}. 
Ideally one might just try to use exactly the same fits and compare it 
to the neutral pions. As the data are from different 
experiments and do not exactly use the same centrality selections, 
the normalization has to be kept as a free parameter. The centrality 
selections used below should however be similar enough, so that the 
shape of the spectra should not differ strongly.

In Figure~\ref{fig:wa98} fits using the parameters obtained above 
are compared to the neutral 
pion spectra for the $12.7 \% $ most central reactions of Pb+Pb at 
158 $A$GeV. The ``best fit'' to NA44 in table~\ref{tbl:parsps} 
is shown as a dashed line, the one to NA49 data as a dotted line. 
The description is only superficially adequate. 
The fit to NA44 overshoots the data at large transverse 
masses, and one obtains a $\chi^{2}/\nu = 712/20$. 
The fit to NA49 is below the data at large transverse 
masses with a $\chi^{2}/\nu = 1395/20$. This large difference for the
pion spectra at high $m_{T}$ is of course due to the much larger flow
velocity for the NA49 fit, which increases the proton inverse slope
while decreasing the one of the pions compared to NA44.
Another fit has been performed keeping only the parameters 
$T_{chem}$ and $\mu_{B}$ the same as in the earlier fit and 
optimizing the kinetic temperature and the flow for the neutral pions. This fit 
(shown as a solid line) yields a good description with $\chi^{2}/\nu = 
27/18$. The fit parameters obtained are
$T_{kin} = 130.3 \, \mathrm{MeV}$ and $\langle \beta_T \rangle = 0.415$. 
The (dis-)agreement of the fits can also be judged from the inset in 
Figure~\ref{fig:wa98}, which shows the ratios of the data to the fits.

\begin{figure}[tb]
        \includegraphics{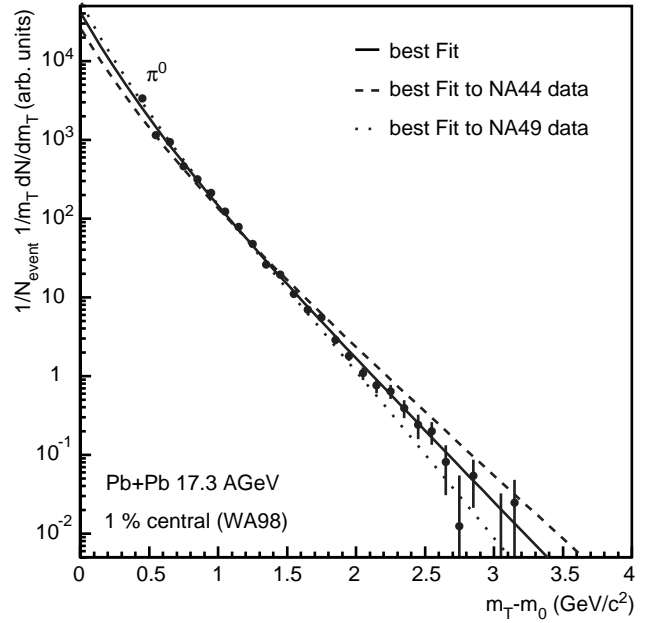}
        \caption{Transverse mass distributions of neutral pions as in 
        Figure~\protect\ref{fig:wa98} for 
        the $1 \%$ most central Pb+Pb collisions at 158 $A$GeV 
        from the WA98 experiment \protect\cite{wa98pi0}.}
        \protect\label{fig:wa98-2}
\end{figure}

One might argue that the different centrality selections are 
responsible for the disagreement. Although this does not seem to be 
very likely, as it has been shown \cite{wa98pi0}, that the  
shape of the pion spectra does not vary strongly for medium-central to 
central collisions, a comparison to the $1 \% $ most central data has 
been performed. The results are displayed in Figure~\ref{fig:wa98-2}. 
Again the dashed  and dotted lines show fits with all parameters fixed as above 
from NA44 and NA49. The behavior of the fits is qualitatively similar
to the one discussed above, the disagreement of the NA49 fit to the
data is not as pronounced as before, which is due to the slightly
steeper spectra in this case.
The solid line shows a good fit with 
$T_{kin} = 113.0 \, \mathrm{MeV}$ and 
$\langle \beta_T \rangle = 0.459$. 
The different parameters for this second fit compared to 
the other central ($12.7 \, \% $) sample are most likely not 
conclusive, but are an example that fitting momentum spectra of just 
one particle species leaves an ambiguity in the parameters. The higher 
flow velocity can be compensated by a lower temperature and vice 
versa. Also I will not put too much emphasis on the fact that the fits 
to the NA44 data appear to overpredict the pion yield at high $p_{T}$, 
because part of this discrepancy might be due to systematic 
differences between the two experiments involved. 

\begin{figure}[tb]
	\includegraphics{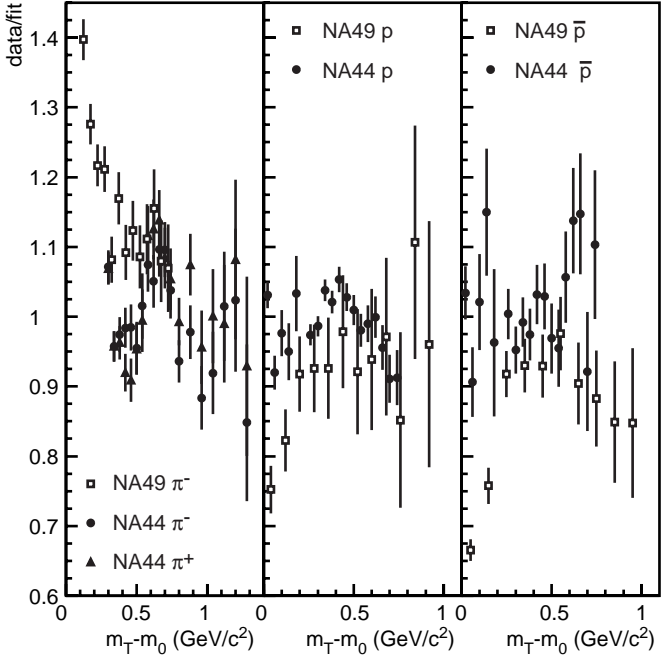}
	\caption{Ratios of experimental transverse mass distributions of pions
	(left), protons (center) and antiprotons (right) to a fit of 
	the hydrodynamical parameterization for 
	central Pb+Pb collisions at 158 $A$GeV 
	from NA44 and NA49.}
	\protect\label{fig:spscomp}
\end{figure}

As one of the major uncertainties in this analysis appears to be
related to different experimental data sets, it may be worthwhile to 
compare data of NA44 and NA49 directly. This has been done in
Fig.~\ref{fig:spscomp} by plotting the ratio of the data to the best 
fit of the parameterization to NA44 data (first row in
table~\ref{tbl:parsps}). The same fit is used for both data sets,
there is one overall normalization for all species which differs
between experiments. One can clearly see that the deviations are
most significant at low $m_{T}$ for protons and antiprotons. 
In the systematic studies it was found that
the parameters for the fits to NA44 do not depend very much on the
fitting range, while the parameters for NA49 change significantly
when the lowest $m_{T}$ data points are not used.
It should also be noted that these fits use data with statistical
errors only, as systematic errors were not readily available for all 
data sets. If systematic errors would account for a significant
fraction of the discrepancy between the two experiments, a
simultaneous fit might be possible and yield more conclusive
information. This could not be attempted in this paper.

One can however note 
that the hydrodynamic parameterization can account for 
a large fraction of the pion production at high $p_{T}$, even if a profile 
with a negligible tail towards higher velocities
($\Delta = 50$ -- similar to the box profile) has been used. A more 
diffuse profile ($\Delta = 5$) would necessarily lead to a larger 
yield at high $p_{T}$. Most of the other systematic variations 
studied do not lead to significant variations in this respect. The
most important uncertainty relates to the discrepancy already seen
between NA44 and NA49. Ignoring the very low $m_{T}$ data points of
NA49 leads to a moderately high flow velocity. In this case the pion 
spectra may be almost entirely be explained by such a source. A very 
high flow velocity as suggested by the low $m_{T}$ NA49 data would considerably
reduce the hydrodynamic yield of pions at high $m_{T}$.

A fixed chemical temperature of $T_{chem} \equiv 165 \, \mathrm{MeV}$ 
leads to different results, but from Fig.~\ref{fig:na44} it is obvious 
that this would also lead to a larger overprediction at high $p_{T}$. 
As sketched above, there may be reasons why this assumption may not 
be adequate for our analysis. In fact, it has been remarked in 
\cite{sollfrank99,kaneta01} that there is a rapidity dependence of 
chemical parameters.
The chemical temperature of
$T_{chem} = 143.9 \, \mathrm{MeV}$ and the baryo-chemical potential of 
$\mu_{B} = 193.3 \, \mathrm{MeV}$ of the best fit can be compared to 
results of other analyses. As stated above, the temperature is lower 
than the values of $T_{chem} = 158 \pm 3 \, \mathrm{MeV}$ \cite{becattini:sps} 
and $T_{chem} = 168 \pm 2.4 \, \mathrm{MeV}$ \cite{pbm99} obtained from 
rapidity integrated yields, but compares well with the result 
 $T_{chem} = 141 \pm 5 \, \mathrm{MeV}$ given in \cite{na44spec}.

\section{Hadron Spectra from RHIC}
The situation at RHIC is more favorable for this analysis, as there 
exist data on different particle species partially reaching out to 
large $p_{T}$ measured under the same conditions within one 
experiment. The PHENIX experiment has presented spectra of neutral 
pions \cite{highpt} and identified charged hadrons 
\cite{ppg006,phenix-hadron} in central Au+Au collisions at 
$\sqrt{s_{NN}} = 130 \, \mathrm{GeV}$. I will also use the data from 
the STAR experiment \cite{starpbar,starp,starpi}, which only cover
low $m_{T}$. From the
STAR experiment, the most recent data on protons and antiprotons use 
a sample of the $6 \% $ most central collisions, while the pions have
been extracted from the $5 \% $ most central collisions. I will
ignore this
difference, as it should not significantly affect the shapes of the spectra,
however, there may be small effects on the relative normalization
which would result in an additional systematic error for the chemical
parameters.

A similar procedure as for the SPS data 
above is followed, with the addition of using $p_{T}$ dependent 
systematic errors. 
For this a systematic error of 
$5 \% $ added in quadrature to the errors for the PHENIX
data (which contain part of the systematic error already) 
has been assumed \cite{ppg009}.
For STAR data a systematic error of $8 \% $ has
been used for the antiproton spectra \cite{starpbar}.  
The proton and pion spectra \cite{starp,starpi} were already quoted
with a systematic error. These does not include all of the
normalization errors, so they might still further affect the chemical
parameters extracted.
Fits are performed to the charged pions for 
$m_{T}-m_{0} < 2 \, \mathrm{GeV}/c$ and to (anti)protons for
$m_{T}-m_{0} < 3 \, \mathrm{GeV}/c$. Systematic variations of the 
fitting assumptions are done as above. In addition I have studied 
fits to a limited transverse mass range 
$m_{T}-m_{0} < 1.5 \, \mathrm{GeV}/c$ for pions and (anti)protons in 
PHENIX.

\begin{figure}[tb]
        \includegraphics{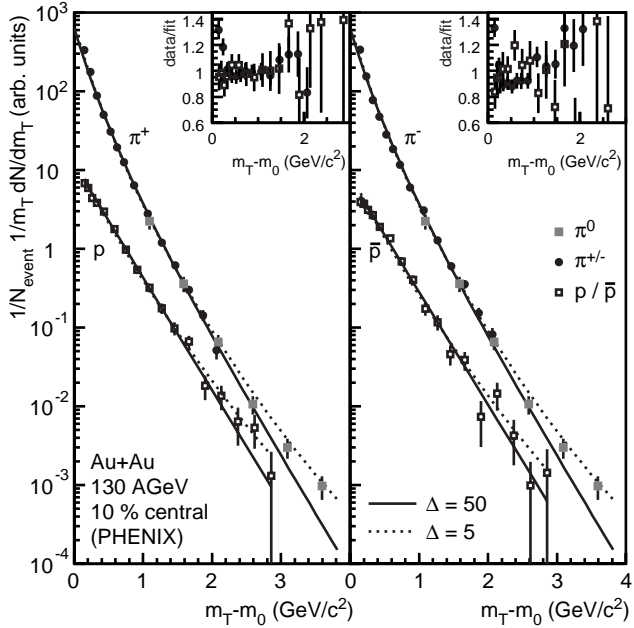}
        \caption{Transverse mass distributions of pions and (anti)protons  for 
        the $10 \%$ most central Au+Au collisions at $\sqrt{s_{NN}} = 
        130 \, \mathrm{GeV}$ 
        from the PHENIX experiment \protect\cite{highpt,ppg006,phenix-hadron}. 
	The lines show fits (A) and (A$^{\prime}$) from table~\ref{tbl:parrhic}.
	The inset shows the ratio of the data to fit (A).}
        \protect\label{fig:phnx-1}
\end{figure}

\begin{figure}[tb]
	\includegraphics{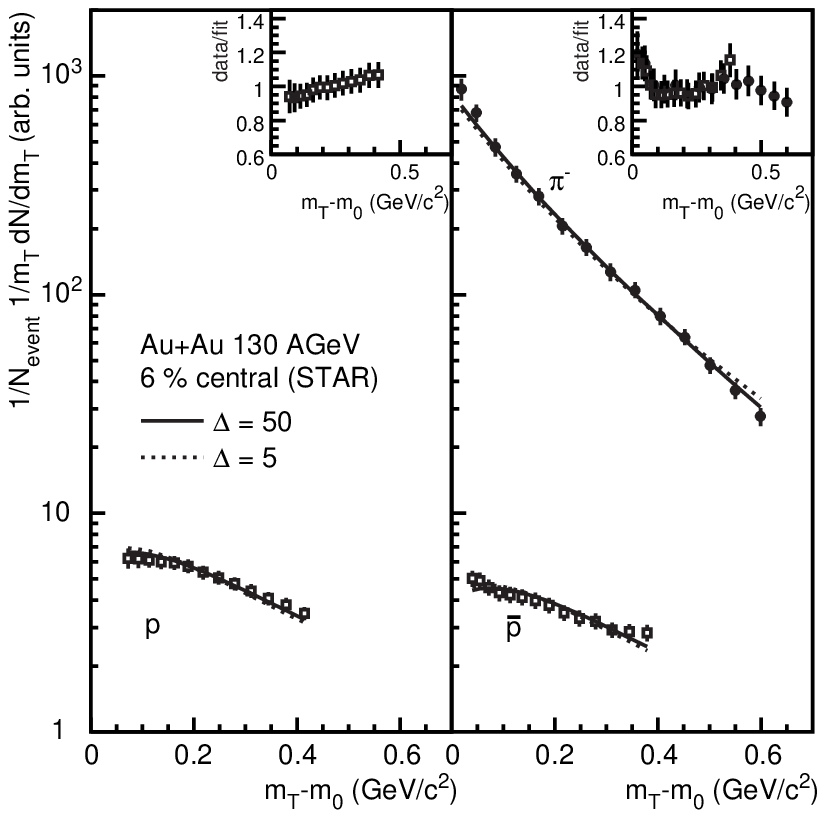}
	\caption{Transverse mass distributions of protons and antiprotons  for 
	the $6 \%$ most central Au+Au collisions at $\sqrt{s_{NN}} = 
	130 \, \mathrm{GeV}$ and for negative pions for 
	the $5 \%$ most central Au+Au collisions
	from the STAR experiment \protect\cite{starpbar,starp,starpi}. 
	The lines show fits (B) and (B$^{\prime}$) from table~\ref{tbl:parrhic}.
	The inset shows the ratio of the data to fit (B).}
	\protect\label{fig:starspec}
\end{figure}

The results are summarized in table~\ref{tbl:parrhic}, a comparison 
of some of the fits to the PHENIX data is shown in Fig.~\ref{fig:phnx-1},
those to the STAR data are displayed in Fig.~\ref{fig:starspec}.

\begin{table*}[tbh]
    \centering
    \caption{Fit parameters of hydrodynamic fits to pion and 
    (anti-)proton spectra in central Au+Au collisions at RHIC. 
    The errors given are fit errors using statistical and partial
    systematic errors (for details see text). No error indicates 
    that this parameter was fixed in the fit.}
    \begin{tabular}{|c||c|c|c|c|c|c|c|}
        \hline
        Remarks & $\chi^{2}/\nu$ & $T_{kin}$ (MeV) & $\eta_{f}$ 
	& $\langle \beta_T \rangle$ & $T_{chem}$ (MeV) 
	& $\mu_{B}$ (MeV) & $\Delta$  \\
        \hline \hline
	\multicolumn 8{|c|}{PHENIX data} \\ 
	\hline
        narrow profile (A) & $101/57$ & $143.0 \pm 6.8$ & $0.745 \pm 
	0.031$ 
	& $0.450  \pm 0.016 $ & $173.3 \pm 2.0$ & $35.9 \pm 4.2$ &
	$50$  \\
        \hline
	wide profile (A$^{\prime}$) & $98/57$ & $142.2 \pm 7.3$ & $0.558 \pm
	0.022$ 
	& $0.414 \pm 0.014$ & $173.0 \pm 2.0$ & $37.1 \pm 4.3$ & $5$  \\
	\hline
	low $m_{T}$ - narrow & $80/39$ & $137.0 \pm 7.4$ & $0.759 \pm 
	0.033$ & $0.457 \pm 0.017$ & $172.5 \pm 2.0$ & $35.0 \pm 4.3$ & 
	$50$  \\
	\hline
	low $m_{T}$ - wide & $72/39$ & $128.5 \pm 8.3$ & $0.609 \pm 
	0.026$ & $0.444 \pm 0.015$ & $172.5 \pm 2.1$ & $35.8 \pm 4.4$ & 
	$5$  \\
	\hline
        no weak decays - narrow & $116/57$ & $144.7 \pm 7.5$ & $0.730 \pm
	0.033$ 
	& $0.442 \pm 0.016$ & $186.4 \pm 2.5$ & $38.6 \pm 4.6$ & $50$  \\
        \hline
	no weak decays - wide & $109/57$ & $141.3 \pm 8.0$ & $0.557 \pm
	0.024$ 
	& $0.413 \pm 0.015$ & $185.5 \pm 2.6$ & $39.9 \pm 4.6$ & $5$  \\
	\hline
	$T_{kin} = T_{chem}$ - narrow & $163/59$ & $165$ & $0.633 \pm 
	0.009$ 
	& $0.392 \pm 0.004$ & $165$ & $43.0 \pm 5.1$ & $50$  \\
	\hline
	$T_{kin} = T_{chem}$ - wide & $158/59$ & $165$ & $0.482 \pm 
	0.006$ 
	& $0.366 \pm 0.004$ & $165$ & $43.8 \pm 5.1$ & $5$  \\
	\hline
	STAR $T$ - narrow & $186/58$ & $94.8$ & $0.962 \pm 
	0.007$ 
	& $0.548 \pm 0.003$ & $163.6 \pm 1.5$ & $35.1 \pm 4.1$ & $50$  \\
	\hline
	STAR $T$ - wide & $206/58$ & $83.6$ & $0.730 \pm 
	0.005$ 
	& $0.509 \pm 0.002$ & $161.4 \pm 1.5$ & $37.5 \pm 4.1$ & $5$  \\
	\hline
	\multicolumn 8{|c|}{STAR data} \\ 
	\hline
	narrow profile (B) & $19/36$ & $94.8 \pm 7.9$ & $1.037 \pm 
	0.031$ 
	& $0.578  \pm 0.012 $ & $165.3 \pm 2.2$ & $31.2 \pm 2.5$ &
	$50$  \\
	\hline
	wide profile (B$^{\prime}$) & $36/36$ & $83.6 \pm 8.3$ & $0.984 \pm
	0.033$ 
	& $0.622 \pm 0.013$ & $174.4 \pm 2.9$ & $33.8 \pm 2.7$ & $5$  \\
	\hline
	no weak decays - narrow & $22/36$ & $88.0 \pm 7.1$ & $1.038 \pm
	0.030$ 
	& $0.578 \pm 0.012$ & $174.6 \pm 2.6$ & $33.0 \pm 2.7$ & $50$  \\
	\hline
	no weak decays - wide & $41/36$ & $77.7 \pm 7.4$ & $0.984 \pm
	0.032$ 
	& $0.622 \pm 0.013$ & $184.9 \pm 3.3$ & $36.0 \pm 2.8$ & $5$  \\
	\hline
	$T_{kin} = T_{chem}$ - narrow & $55/38$ & $165$ & $0.920 \pm 
	0.020$ 
	& $0.530 \pm 0.008$ & $165$ & $32.4 \pm 2.5$ & $50$  \\
	\hline
	$T_{kin} = T_{chem}$ - wide & $75/38$ & $165$ & $0.781 \pm 
	0.019$ 
	& $0.535 \pm 0.009$ & $165$ & $34.0 \pm 2.5$ & $5$  \\
	\hline
	PHENIX $T$ - narrow & $41/37$ & $143.0$ & $0.944 \pm 
	0.029$ 
	& $0.540 \pm 0.012$ & $164.4 \pm 2.4$ & $32.1 \pm 2.5$ & $50$  \\
	\hline
	PHENIX $T$ - wide & $61/37$ & $142.2$ & $0.858 \pm 
	0.032$ 
	& $0.570 \pm 0.014$ & $170.4 \pm 3.2$ & $34.2 \pm 2.6$ & $5$  \\
	\hline
	\multicolumn 8{|c|}{PHENIX + STAR data} \\ 
	\hline
	narrow profile (C) & $206/98$ & $112.3 \pm 3.3$ & $0.886 \pm 
	0.014$ 
	& $0.515  \pm 0.006 $ & $165.0 \pm 1.0$ & $34.5 \pm 2.2$ &
	$50$  \\
	\hline
	wide profile & $323/98$ & $108.3 \pm 4.0$ & $0.673 \pm
	0.012$ 
	& $0.479 \pm 0.006$ & $160.9 \pm 1.0$ & $38.1 \pm 2.1$ & $5$  \\
	\hline
	no weak decays - narrow & $230/98$ & $109.4 \pm 3.5$ & $0.888 \pm
	0.015$ 
	& $0.516 \pm 0.006$ & $175.5 \pm 1.3$ & $36.7 \pm 2.3$ & $50$  \\
	\hline
	no weak decays - wide & $332/98$ & $103.8 \pm 4.2$ & $0.680 \pm
	0.012$ 
	& $0.483 \pm 0.006$ & $170.4 \pm 1.2$ & $40.3 \pm 2.3$ & $5$  \\
	\hline
	$T_{kin} = T_{chem}$ - narrow & $389/100$ & $165$ & $0.672 \pm 
	0.006$ 
	& $0.413 \pm 0.003$ & $165$ & $38.5 \pm 2.1$ & $50$  \\
	\hline
	$T_{kin} = T_{chem}$ - wide & $470/100$ & $165$ & $0.517 \pm 
	0.004$ 
	& $0.388 \pm 0.002$ & $165$ & $40.1 \pm 2.1$ & $5$  \\
	\hline
   \end{tabular}
    \label{tbl:parrhic}
\end{table*}

The fits to PHENIX data show slightly better agreement for a wide
profile ($\Delta = 5$), however, the difference in $\chi^{2}$ is
small. As can be seen in Fig.~\ref{fig:phnx-1}, the fit for $\Delta = 5$
(dotted line)
has a more important contribution at high $p_{T}$.
The STAR data are generally better described using the narrow 
profile, the differences are, however, not easily discernible 
in Fig.~\ref{fig:starspec}. For PHENIX, the two different kinds of
fits yield very similar fit parameters, while there are slightly
stronger variations when looking at the STAR data. As there does not 
seem to be a strong preference for the wider profile in any of the
data sets and as there are also theoretical arguments from true
hydrodynamical calculations \cite{huovinen01}, I will base most 
of the discussion below on the fits with a narrow profile.
For both data sets fits ignoring weak decays yield higher
chemical temperatures and slightly lower kinetic temperatures. The
biggest difference is again between the two experiments. The
chemical parameters still show only a small effect, e.g. while fit
(A) to PHENIX data yields 
$T_{chem} = 173.3 \, \mathrm{MeV}$ and
$\mu_{B} = 35.9 \, \mathrm{MeV}$, the values for STAR (B) are slightly
smaller with $T_{chem} = 165.3 \, \mathrm{MeV}$ and
$\mu_{B} = 31.2 \, \mathrm{MeV}$. These values are similar 
to the results in \cite{pbm-rhic} where
$T_{chem} = 174\pm 7  \, \mathrm{MeV}$ and
$\mu_{B} = 46\pm 5 \, \mathrm{MeV}$ are given. The better agreement 
at RHIC compared to SPS of 
the chemical parameters of this analysis with those of the integrated 
yields may be related to the fact that boost 
invariance is a much better approximation in the RHIC case.
Larger differences are seen for the
kinetic parameters. Fit (A) yields 
$T_{kin} = 143.0 \, \mathrm{MeV}$ and
$\langle \beta_T \rangle = 0.450$, fit (B) 
$T_{kin} = 94.8 \, \mathrm{MeV}$ and
$\langle \beta_T \rangle = 0.578$. As a check also fits have been
performed which use the kinetic temperature from one experiment as a
fixed parameter in a fit to the other experiment, and vice versa. 
A reasonable fit to
the STAR data (with a higher $\chi^{2}$) 
can be obtained this way by using a high kinetic
temperature, but the flow velocity turns out to be larger than for
PHENIX. In the other case, the fit to PHENIX data with a low
temperature yields a much higher $\chi^{2}$ -- no good fit can be
obtained.
There have been fits to the PHENIX data using hydrodynamic
parameterizations that do not include resonances \cite{qm02:jane}.
These yield similar average 
flow ($\langle \beta_T \rangle = 0.47 \pm 0.01$)
velocities but smaller kinetic
freeze-out temperatures ($T_{kin} = 121 \pm 4 \, \mathrm{MeV}$) -- this 
difference is consistent with the
influence of resonance decays.

Fits assuming a unique temperature of 
$T = 165 \, \mathrm{MeV}$ for both kinetic and chemical freeze-out as in 
\cite{broniowski} are again much worse in fit quality for both shapes
of the source distribution, as can be seen from the large 
$\chi^{2}$ in table~\ref{tbl:parrhic}. The 
assumption of simultaneous kinetic and chemical freeze-out underlying this 
fit is questionable, but I will still consider such a fit once more
below. 
One might argue that the transverse mass range used in these fits 
does include possible contributions of hard scattering. However, fits which use 
data only for low $m_{T}$ ($m_{T}-m_{0} < 1.5 \, \mathrm{GeV}/c$) 
yield very similar 
results as seen in table~\ref{tbl:parrhic}.

\begin{figure}[tb]
	\includegraphics{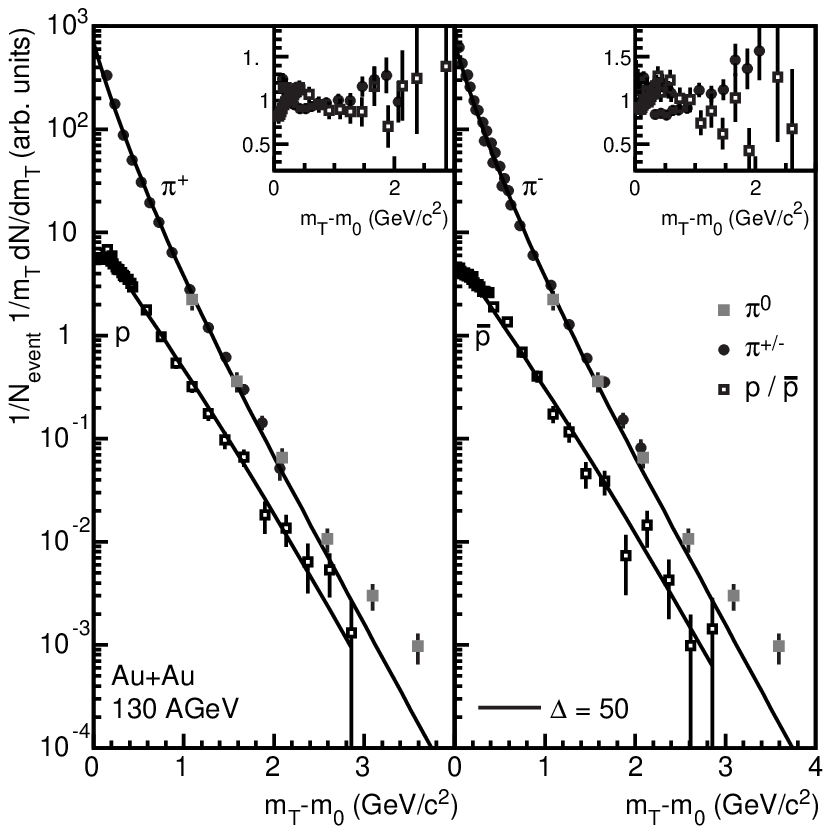}
	\caption{Transverse mass distributions of pions and (anti)protons  for 
	central Au+Au collisions at $\sqrt{s_{NN}} = 
	130 \, \mathrm{GeV}$ 
	from the PHENIX and STAR experiments. 
	The lines show a simultaneous fit (C) from table~\ref{tbl:parrhic}.
	The inset shows the ratio of the data to the fit.}
	\protect\label{fig:rhicall}
\end{figure}

Finally, a simultaneous fit to STAR and PHENIX data has been
performed, where 
the STAR data have been rescaled by a factor 0.92 (ratio
of the number of participants) to account for the different
centrality selection. 
Please note that such a simultaneous fit is only
successful if one includes the systematic errors. Considering just statistical 
errors, the data from the two experiments could not be consistently
described by this model. 
The results are also given in table~\ref{tbl:parrhic}, the best fit
(C) is
compared to the data in Fig.~\ref{fig:rhicall}. Fit (C) is obtained
using $\Delta = 50$ and describes the data reasonably well with 
$\chi^{2}/\nu = 206/98$. Chemical parameters change relatively 
little. As might be expected the kinetic parameters are intermediate 
between those for the individual experiments, one finds 
$T_{kin} = 112.3 \, \mathrm{MeV}$ and
$\langle \beta_T \rangle = 0.515$. Once again a fit using 
$T_{kin} = T_{chem} = 165 \, \mathrm{MeV}$ is significantly worse.

\section{Discussion}

The results for SPS and RHIC data may now be compared to each other. 
In both cases a
box-like profile is favored over a
Gaussian-like profile. It was already discussed in \cite{wa98:hydro:99} that a 
box-like profile 
would be required in case of a large flow velocity, and this was also found to 
be the preferred profile in \cite{tomasik99}. 
Such a distribution with little or no ``tail'' to higher velocities
is also closer to expectations from true hydrodynamic calculations
\cite{huovinen01}.
In both cases (SPS and RHIC) there are fits with a range of flow
velocities, depending mainly on the experiment that one concentrates 
on. 
At the SPS the range of velocities is $\langle \beta_T \rangle = 0.48
- 0.55$ and the analysis presented here does not allow to constrain
this value further. At RHIC a combined fit is possible, which yields 
as the most likely value $\langle \beta_T \rangle = 0.515$.
Similarly the range of kinetic freeze-out temperatures for the SPS is 
$T_{kin} = 76 - 122 \, \mathrm{MeV}$, while the RHIC value appears to
be $T_{kin} = 112.3 \, \mathrm{MeV}$. 
Apparently this analysis indicates that the flow velocities at SPS
and RHIC may be similar, and there is no clear indication of
stronger collective flow at RHIC as found in \cite{xu:qm01}. It can, 
however, be seen in \cite{xu:qm01} that this is strongly influenced by 
the proton and antiproton spectra from STAR used in the analysis, 
which show a much larger inverse slope than the PHENIX results.

Both for SPS and RHIC there are still considerable uncertainties
which are related to the experimental data themselves. For the SPS
these imply large uncertainties for the extrapolation of the
hydrodynamical fits to higher $m_{T}$.
Still, if we attempt to ascribe particle production at low $m_{T}$ 
in heavy-ion collisions at the 
SPS to hydrodynamic scenarios, the 
extrapolation of the hydrodynamic contribution to high $m_{T}$ will 
yield a considerable contribution of hydrodynamic production. 

\begin{figure}[tb]
        \includegraphics{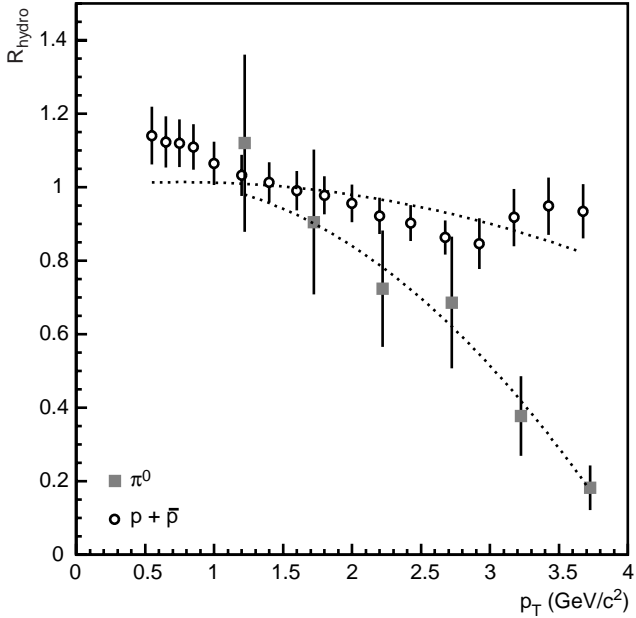}
        \caption{Fraction of the hydrodynamic component of the 
        neutral pion and proton+antiproton yield using fit C 
        from table~\ref{tbl:parrhic}. The dotted line 
	shows a fit of a Woods-Saxon function to the data (see 
        text).}
        \protect\label{fig:ratio}
\end{figure}

For RHIC the measurement of protons and antiprotons at higher $m_{T}$
does provide additional constraints and excludes some of the extreme 
scenarios. A simultaneous fit to STAR and PHENIX has been performed,
which yields a reasonable agreement with the available data.
In the following I will assume that this fit describes the
contribution of a hydrodynamic source to particle emission. It does 
describe the spectra almost entirely at low $m_{T}$. At higher
$m_{T}$ the fit starts to deviate, which is not
surprising, as one would expect additional production mechanisms,
especially hard scattering, to play a role there. However, this
hydrodynamic contribution may be still significant even at high
momenta -- this would be of particular relevance to any study of hard
scattering, because to be able to obtain information on hard
scattering production one would have to know this hydrodynamic
contribution and take it into account. 

A simple partitioning of the produced particles into either
hydrodynamic or hard scattering production is not possible. If one
believes that equilibration is taking place in these collisions,
there will almost necessarily be parts of the reaction volume that
are only partially equilibrated. At RHIC the high $p_{T}$ part of the
non-equilibrated spectrum (with a power-law shape from hard
scattering) should have a larger inverse slope than the equilibrated 
spectrum (similar to exponential). So in this case equilibration
through multiple collisions will slow particles down and soften the
spectrum. As the degree of equilibration will vary throughout the
reaction volume, the spectrum might be a superposition of a range of 
spectra between the one from the initial collisions and a
thermalized one.

At this point one does not necessarily need to distinguish between
parton and hadron degrees of freedom. Thus, parton energy loss would 
just be one possible equilibration mechanism. While it is not
possible to precisely predict how close the final distribution would 
be to a thermal (hydrodynamic) distribution, the latter can be taken 
as a limiting case. If hard scattered partons are completely
quenched, one will still end up with a thermal distribution as a
minimum yield. Any excess over this minimum might be due to
non-equilibrated contributions, either true hard scattering
production or not completely equilibrated production, which still has
a higher average momentum than the thermal.

The discussion in this paper assumes that the spectra in the
intermediate momentum region can be described by thermal
distributions. This is not completely correct, as e.g. even hard
scatterings will produce some lower momentum component. However, as
any comparison of pure hydrodynamics to data, I will assume for the
following that this non-thermal contribution at low and intermediate 
$p_{T}$ is negligible.

In Fig.~\ref{fig:ratio} ratios of the simultaneous fit (C) 
to the experimental data, 
i.e. the fractions $R_{hydro}$ of the produced particles accounted for by the 
hydrodynamic parameterization, are shown for neutral pions (solid squares) and 
for protons+antiprotons (open circles) as a function of $p_{T}$. 

The proton+antiproton yield is reasonably described by this fit in
the entire $p_{T}$ range investigated, as can be seen by the more or 
less constant ratio. The hydrodynamic contribution to 
pions depends more strongly on $p_{T}$. While here 
hydrodynamics dominates also at lower $p_{T}$, there is a steady 
decrease of the contribution towards higher $p_{T}$. Still even for
$p_{T} \ge 3 \, \mathrm{GeV}/c$ the hydrodynamic contribution to the pion 
yield is close to $40 \%$.

The not completely thermalized part of the emitted hadrons will
reflect itself in other observables. A $p_{T}$ dependent non-thermalized
contribution to the spectrum will be related to a $p_{T}$ dependent
deviation of the measured elliptic flow $v_{2}$ from the purely
hydrodynamic behavior, because the latter relies on equilibration
achieved earlier in the collision history than kinetic freeze-out.
Quantitative estimates of this are difficult, as the partially
thermalized contribution will lead to an unknown contribution to
$v_{2}$. Full hydrodynamic calculations, when applied to elliptic
flow and particle spectra, should be able to extract more information
on this.

\section{Summary}

Results of hydrodynamic fits to momentum spectra at mid-rapidity 
in heavy ion reactions at the SPS and at RHIC have been presented.
While at RHIC a simultaneous description of STAR and PHENIX data is
possible, when systematic errors are taken into account, there are
still unresolved differences at SPS depending on the experiment used. 
At SPS a chemical temperature of 
$T_{chem} \approx 145 \, \mathrm{MeV}$ and a baryo-chemical potential of 
$\mu_{B} \approx 190 \, \mathrm{MeV}$ have been obtained. This
chemical temperature is slightly smaller than values obtained from
rapidity integrated yields. 
At RHIC the corresponding values are
$T_{chem} \approx 165 \, \mathrm{MeV}$ and  
$\mu_{B} \approx 35 \, \mathrm{MeV}$. The kinetic parameters are
$T_{kin} = 76 - 122 \, \mathrm{MeV}$ and 
$\langle \beta_T \rangle = 0.48
- 0.55$ at SPS and
$T_{kin} \approx 110 \, \mathrm{MeV}$ and 
$\langle \beta_T \rangle \approx 0.52$ at RHIC when using a 
velocity profile similar to a box profile. 
The momentum spectra at RHIC are broader than at 
the SPS. However, this analysis does not prove a significantly 
stronger transverse flow at RHIC, the change in the momentum spectra 
might also be due to a higher freeze-out temperature.
At high transverse momenta the spectra of 
protons can be nicely described at RHIC. 
The pion spectra leave room for additional 
particle production mechanisms (e.g. hard scattering) at high $p_{T}$. 
Estimates of the suppression of hard 
scattering should, however, take the non-negligible contribution from 
a hydrodynamic source into account. 
Important hydrodynamic contributions 
in spectra at high $p_{T}$ require even stronger quenching of hard 
scattering production. In return, quenching is a necessary condition 
to achieve locally equilibrated distributions.

\begin{acknowledgement}
\begin{sloppypar}
The author would like to thank U.A.~Wiedemann for helpful discussions 
and J.M.~Burward-Hoy, M.~Calder\'{o}n~de~la~Barca~S\'{a}nchez, 
M.~Gazdzicki, M.~Kaneta, K.~Schweda, R.~Snellings
for valuable information regarding the experimental data. Part of
this work was created at the University of M{\"u}nster and at GSI
Darmstadt.
\end{sloppypar}
\end{acknowledgement}

\end{document}